\documentclass[12pt,preprint]{aastex}
\usepackage{spr-astr-addons}

\newcommand{\HII}{H\,{\sc ii}~}

\newcommand{\Msun}{\rm M_\odot}

\newcommand{\Msunpyr}{\ensuremath{{\rm M}_{\odot}{\rm yr}^{-1}}}
\newcommand{\mum}{\ensuremath{\mu\mbox{m}}}

\begin{document}

\title{Spaxel Analysis: Probing the Physics of Star Formation in Ultraluminous Infrared Galaxies}
\shorttitle{Star Formation in ULIRGs}
\shortauthors{Dopita et al.}

\author{ Michael A. Dopita\altaffilmark{1}\altaffilmark{2}\altaffilmark{3},  Jeffrey Rich \altaffilmark{4}, Fr\'ed\'eric P.~A. Vogt \altaffilmark{1}, Lisa J. Kewley\altaffilmark{1}\altaffilmark{3}, I-Ting Ho\altaffilmark{3}, Hassan M. Basurah \altaffilmark{2}, Alaa Ali\altaffilmark{2} \& Morsi A. Amer\altaffilmark{2}}
\email{Michael.Dopita@anu.edu.au}
\altaffiltext{1}{Research School of Astronomy and Astrophysics, Australian National University, Cotter Rd., Weston ACT 2611, Australia }
\altaffiltext{2}{Astronomy Department, King Abdulaziz University, P.O. Box 80203, Jeddah, Saudi Arabia}
\altaffiltext{3}{Institute for Astronomy, University of Hawaii, 2680 Woodlawn Drive, Honolulu, HI 96822, USA}
\altaffiltext{4}{Carnegie Observatories, 813 Santa Barbara Street, Pasadena, CA 91101, USA}

\email{Michael.Dopita@anu.edu.au}
\begin{abstract}
This paper presents a detailed spectral pixel (spaxel) analysis of the ten Luminous Infrared Galaxies (LIRGs) previously observed with the Wide Field Spectrograph (WiFeS), an integral field spectrograph mounted on the ANU 2.3m telescope, and for which an abundance gradient analysis has already been presented by \citet{Rich12}. Here we use the strong emission line analysis techniques developed by \citet{Dopita13} to measure the ionisation parameter and the oxygen abundance in each spaxel. In addition, we use the observed H$\alpha$ flux to determine the surface rate of star formation (\Msunpyr kpc$^{-2}$) and use the [\ion{S}{2}] $\lambda\lambda6717/6731$ ratio to estimate the local pressure in the ionised plasma. We discuss the correlations discovered between these physical quantities, and use them to infer aspects of the physics of star formation in these extreme star forming environments. In particular, we find a correlation between the star formation rate and the inferred ionisation parameter. We examine the possible reasons for this correlation, and determine that the most likely explanation is that the more active star forming regions have a different distribution of molecular gas which favour higher ionisation parameters in the ionised plasma.\end{abstract}

\section{Introduction}
The Ultraluminous Infrared Galaxies (ULIRGs) provide a vital link between star formation processes occurring in the local universe, and star formation in high redshift galaxies. Although rather rare in the local Universe, ULIRGs account for a much greater fraction of star formation by $z\sim 1$ when encounters between galaxies was much more frequent \citep{Le Floc'h05,Magnelli11}. By the study of ULIRGs, we hope to understand the physics triggering star formation in merging systems, and to discover whether, for example, the initial mass function (IMF) is biased towards more massive stars \citep{Hoversten08, vanDokkum08,Meurer09,Treu10, Cappellari12,Andrews13, Bekki13}, or discover whether the mass distribution of stellar clusters themselves is somehow different in regions of very high star formation rates \citep{Portegies10,Chandar10,Adamo10,Weidner11}. Although these issues have been studied through surface photometry, the advent of integral field spectrographs is relatively recent, and the use of these to simultaneously measure the relationship of star formation, metallicity, gas pressure and the excitation of the \HII regions surrounding the newly formed star clusters is only in its infancy.

In this paper, we investigate the utility of spaxel (spectral pixel) analysis for the study of local ULIRGs which are not dominated by shock-excited gas such as those described by \citet{Rich11}. The targets are ULIRGs drawn from the Great Observatory All-Sky LIRG Survey (GOALS) sample \citep{Armus09}. GOALS is a multi-wavelength survey of the brightest 60\mum\ extragalactic sources in the local universe ($\log(L_{\rm IR}/L_{\odot}) > 11.0$) and with redshifts $z < 0.088$. The GOALS sample is a complete subset of the IRAS Revised Bright Galaxy Sample (RBGS) \citep{Sanders03}. Objects in GOALS cover the full range of nuclear spectral types and interaction stages and, given their luminosity distribution, serve as useful analogs for comparison with high-redshift galaxies.

\section{Observations}
The galaxies analysed here are the same as those analysed by \citet{Rich12} for the derivation of the abundance gradients and reddening. We refer the reader to that paper for the spatial orientations, reddening maps and other global properties of these galaxies. All of the galaxies lie in the restricted luminosity range $11.93 > \log(L_{\rm IR}/L_{\odot}) > 11.41$. Table \ref{Table_1} gives the names, the adopted luminosity distances and the IR luminosities of the target objects.

All observations were made with the Wide Field Field Spectrograph (WiFeS) located at the Mount Stromlo and Siding Spring Observatory 2.3m telescope. This dual-beam image-slicing integral field spectrograph is described by \citet{Dopita07}, and its on-telescope performance is described in \citet{Dopita10}. It provides a $25 \times 38$ arc sec. field of view with the 1.0 arc sec. square spaxels fully filling the field. Our observations used the R7000 and B3000 gratings giving a 5700-7000\AA\ spectral range in the red at a resolution of $R \sim 7000$ (FWHM = 40 km s$^{-1}$) and spectral coverage of 3700-5700\AA\ at a resolution of $R \sim 3000$ (FWHM = 100 km s$^{-1}$) in the blue. The data reduction and analysis procedures are fully described in earlier papers \citep{Rich10,Rich11,Rich12} and will not be repeated here.
\begin{table}
\caption{The GOALS galaxies analysed in this paper.
     }\label{Table_1}
\centering
\resizebox{!}{3.0cm} {\begin{tabular}{llcc}
\noalign{\smallskip} \hline\hline\noalign{\smallskip}
 (1) & (2) & (3) & (4) \\
IRAS No. &  Other Name & Distance & $\log(L_{\rm IR}/L_{\odot})$ \\
  &  & (Mpc) &  \\
 \noalign{\smallskip} \hline \noalign{\smallskip} \\
F01053-1746 & IC 1623 A/B & 82.3 &11.71\\
IRAS 08355-4944 & -- & 112 & 11.62\\
F10038-3338 & ESO 374-IG032 & 145.8 & 11.78 \\
F10257-4339 & NGC 3256 & 38.2 & 11.64 \\
F13373+0105 E/W & Arp 240 & 103.5 & 11.62  \\
F17222-5953 & ESO 138-G027 & 94 & 11.41\\
F18093-5744 N/S & IC 4687/4689 & 79/82 & 11.62  \\
F18341-5732 & IC 4734 & 70.9 & 11.35 \\
F19115-2124 & ESO 593-IG008 & 201 & 11.93\\
 \noalign{\smallskip} \hline \noalign{\smallskip} \\
 
\end{tabular}}
\end{table}

\section{The Spaxel Analysis}
In our analysis we seek to attach  to each spectral pixel (spaxel) a star formation rate, a gas pressure, $P/k$, an oxygen abundance, $12+\log$(O/H), and an ionisation parameter $q$ (related to the dimensionless ionisation parameter ${\cal U} = q/c$, where $c$ is the speed of light). For each spaxel, de-reddened emission line fluxes were derived for all the strong lines from the [\ion{O}{2}] doublet $\lambda\lambda$ 3726,29 all the way up to the [\ion{S}{2}] doublet $\lambda\lambda$ 6717,31. 

The star formation is derived from the H$\alpha$ flux. This is possible because flux in a hydrogen line is proportional to the number
of photons produced by the hot stars, which is in turn proportional to their birthrate. This relationship was earlier calibrated by \citet{Dopita94}. Here we use the ``standard'' calibration for starbursts by \citet{Kennicutt98}, based on the Starburst 99 models of \citet{Leitherer95}: 
\begin{equation}
SFR_{{\rm H\alpha }}=7.9\times 10^{-42}\left[ L_{{\rm H\alpha }}/{\rm
erg.s}^{-1}\right]  \label{eqn1}
\end{equation}
Using the surface brightness per square arc sec. and the distance we can translate this star formation rate to the (local) surface rate of star formation in units of \Msunpyr kpc$^{-2}$.

The mean gas pressure in the ionised plasma is derived from the [\ion{S}{2}] $\lambda\lambda 6731/6717$ ratio. This line ratio is almost insensitive to electron density below $n_e \sim 50$ cm$^{-3}$, so in effect pressures of $log(P/k) < 5.5$ Kcm$^{-3}$ cannot be measured. Since most of the galaxies measured here have super-solar metallicities, their electron temperatures are low; $T_e \lesssim 5000$ K. In addition, the electron densities and electron temperatures of such \HII regions are computed to vary strongly throughout their volume. To account for this we ran a supplementary set of \emph{Mappings IV} models as described in \citet{Dopita13} for abundances of 2, 3 and 5$Z_{\odot}$ ($12+\log(\rm{O/H})=8.99, 9.17$ and 9.39 respectively) and with a set of fixed pressures in the range $5.0 \leq \log(P/k) \leq 7.0$ Kcm$^{-3}$. This provided a theoretical relationship between $\log(P/k)$ and the actual [\ion{S}{2}] $\lambda\lambda 6731/6717$ ratio. This was used to transform the  measured [\ion{S}{2}] line ratio to a local pressure in each spaxel.

From the line flux data we prepared two abundance-sensitive line ratios described in \citet{Dopita13};  
\begin{itemize}
\item{ [\ion{N}{2}]$\lambda6584$/[\ion{O}{2}]$\lambda\lambda3727,9$; and }
\item {[\ion{N}{2}]$\lambda6584$/[\ion{S}{2}]$\lambda\lambda6717,31$, }
\end {itemize}
and three excitation-dependent ratios:
\begin{itemize}
\item {[\ion{O}{3}]$\lambda5007$/[\ion{O}{2}$]\lambda\lambda3727,9$, }
\item {[\ion{O}{3}]$\lambda5007$/[\ion{S}{2}]$\lambda\lambda6717,31$ and }
\item {[\ion{O}{3}]$\lambda5007$/H$\beta$.}
\end {itemize}

This gives six possible diagnostic plots which are fed through the {\sf pyqz} Python module described in \citet{Dopita13} to return the computed value of $\log(q)$ and $12+\log$(O/H) for each of the six pairs of line ratios, as well as the mean and standard deviation computed from these.
Our {\sf pyqz} Python module (v0.4 is freely available for the community to use under the GNU General Public License (doi:10.4225/13/516366F6F24ED). \footnote{ A digital object identifier (DOI) is a string of character used to uniquely identify an object such as an electronic document. The storage location (a.k.a url) corresponding to a given DOI can be resolved using dedicated websites, such as http://dx.doi.org (this tool also allows direct access at http://dx.doi.org/10.4225/13/516366F6F24ED)}.

\begin{figure}[ht]
\includegraphics[width=0.9\hsize]{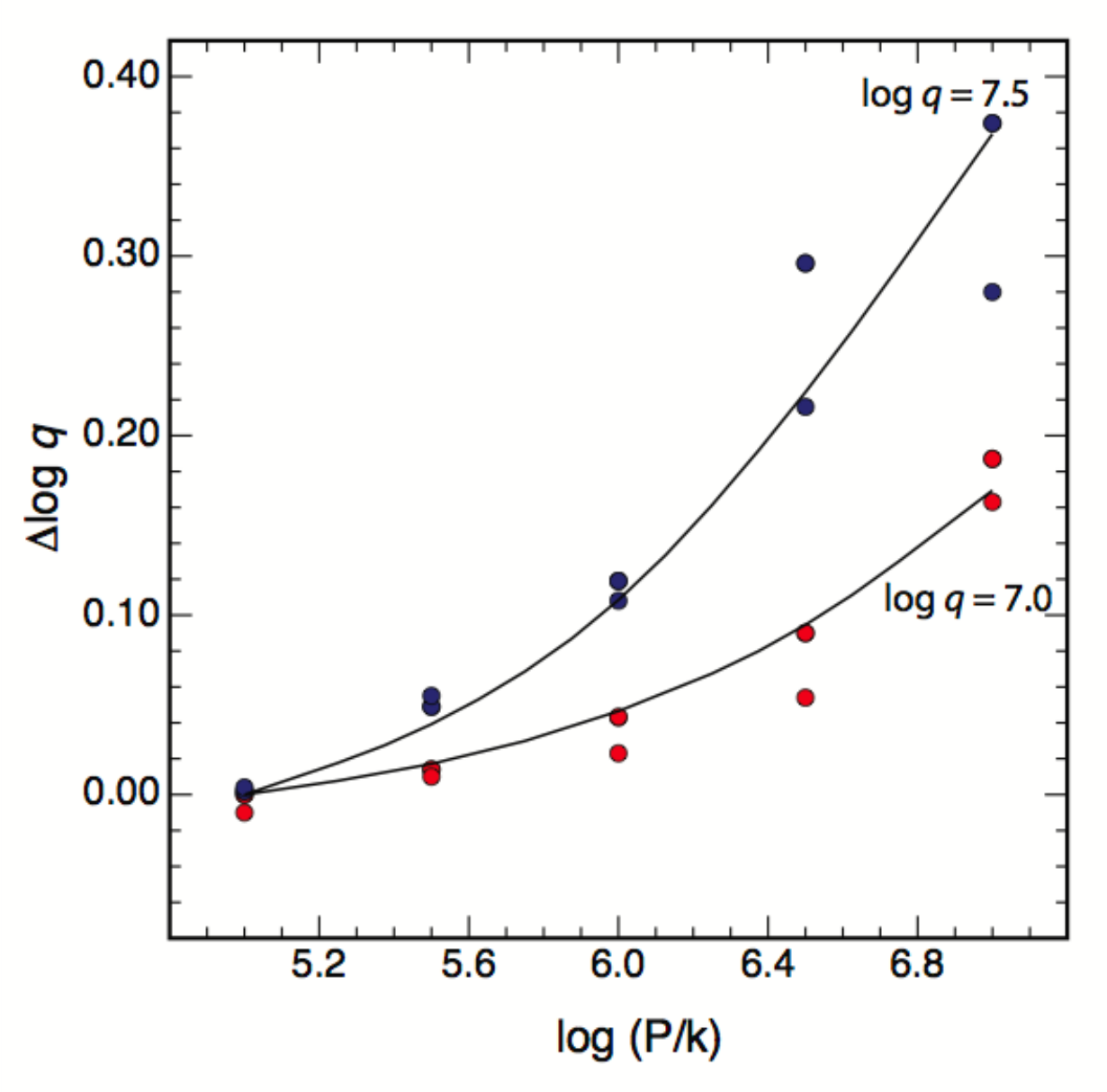}
\caption{The overestimate in the logarithmic ionisation parameter $\Delta \log q$ computed as a function of pressure $\log(P/k)$ for three separate abundances (here, the 2 and 3 times solar abundance models overlap). The curves give the empirical fitting formula we have derived, given in equation \ref{eqn_2}. }\label{fig_1}
\end{figure}

A complication of this analysis is that the line ratios are a function of pressure in the \HII region. The effect of this is mostly to change [\ion{O}{3}]$\lambda5007$/H$\beta$, although other line ratios are affected to lesser degree. As a result, the ionisation parameter derived in high pressure regions may be systematically overestimated. To account for this effect, we ran additional  \emph{Mappings IV}  models as described by \citet{Dopita13}, for a set of fixed ionisation parameters, $12+\log(\rm{O/H})=8.99, 9.17$ and 9.39 respectively and with a set of (high) pressures in the range $5.0 \leq \log(P/k) \leq 7.0$ Kcm$^{-3}$. The resulting six pairs of line ratios were input to the {\sf pyqz} Python module (v0.4) to solve for both $\log(q)$ and $12+\log(\rm{O/H})$, and the difference between the values returned and those of the input model were determined. For the abundances, the error was everywhere less than 0.012 dex. However, the error in $\log(q)$, $\Delta \log q$ was much greater. This is displayed in Figure \ref{fig_1}. On the basis of these models we have determined an empirical fit to $\Delta \log q$ which provides a satisfactory approximation over the full range of parameters;
\begin{equation}
\Delta \log q = 0.04(\log q-6.6)\left[x+1.5x^{1.8}\right] \label{eqn_2}
\end{equation}
where $q$ is the uncorrected value returned by the {\sf pyqz} Python module, and $x = \log(P/k) -5$. The pressure is determined from the  [\ion{S}{2}] $\lambda\lambda 6731/6717$ ratio, as described above. The correction of equation \ref{eqn_2} was applied to each spaxel for which the pressure could be measured.

 \begin{figure}[ht]
\includegraphics[width=1.0\hsize]{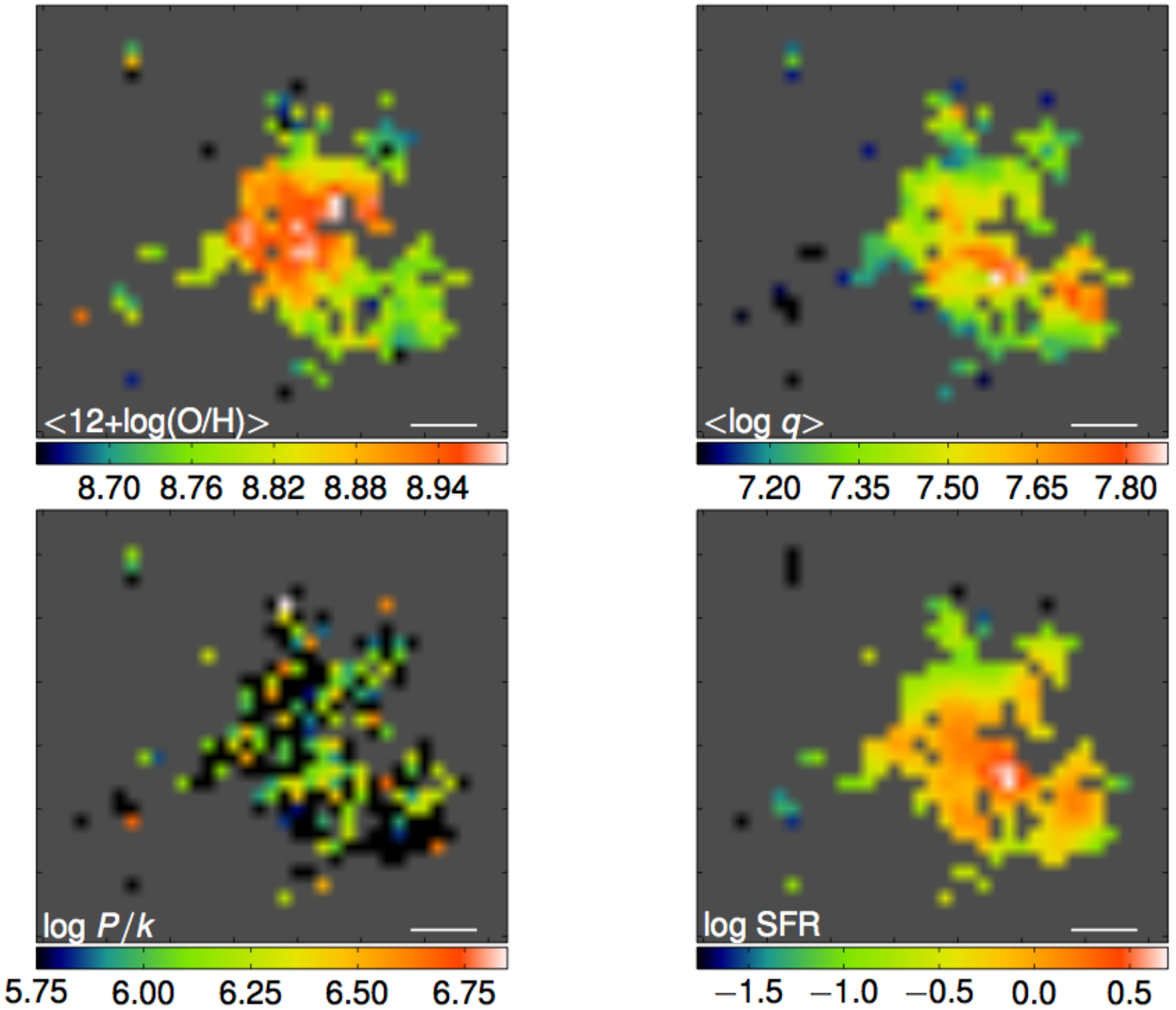}
\caption{The spaxel analysis maps of F01053-1746 (IC1623 A/B). Top left: the abundance gradient map, 12+$\log$(O/H). Top right: the ionisation parameter, $\log(q)$. Bottom left: the local pressure in the ionised plasma, $\log(P/k)$ (pressures below $\log(P/k) \sim 5.7$ cannot be inferred from the [\ion{S}{2}] $\lambda\lambda6717/6731$ ratio). Bottom right: the star formation rate $\log(\rm SFR)$ in \Msunpyr kpc$^{-2}$. The short horizontal white bar represents 5 arc sec. on the sky. Note that the correlation between  $\log(\rm SFR)$ and $\log(q)$ is better than between  $\log(\rm SFR)$ and 12+$\log$(O/H). }\label{fig_2}
\end{figure}

 \begin{figure}[ht]
 \includegraphics[width=1.0\hsize]{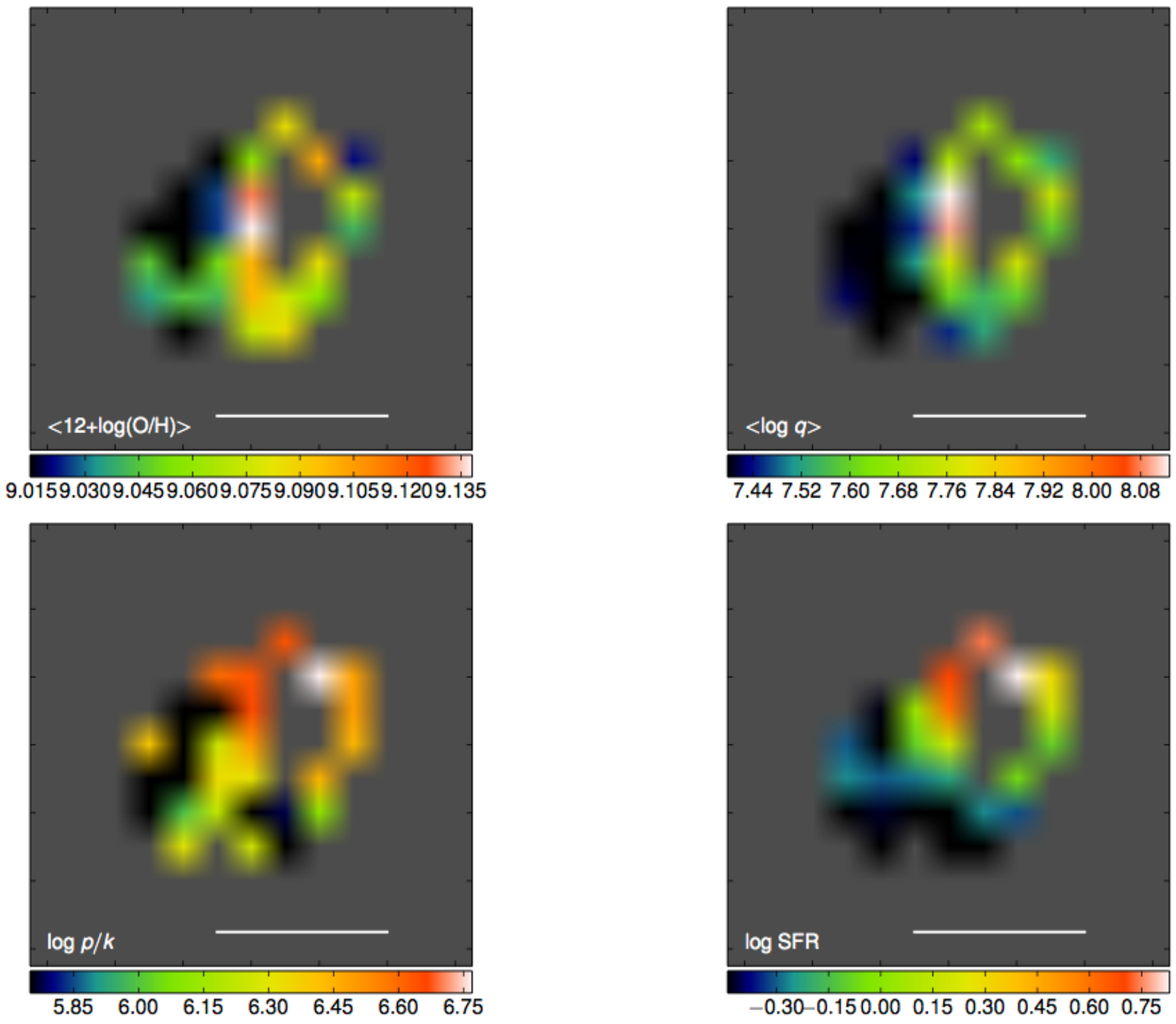}
\caption{As figure \ref{fig_1} but for the galaxy  IRAS08355-4944.}\label{fig_3}
\end{figure}

 \begin{figure}[ht]
\includegraphics[width=1.0\hsize]{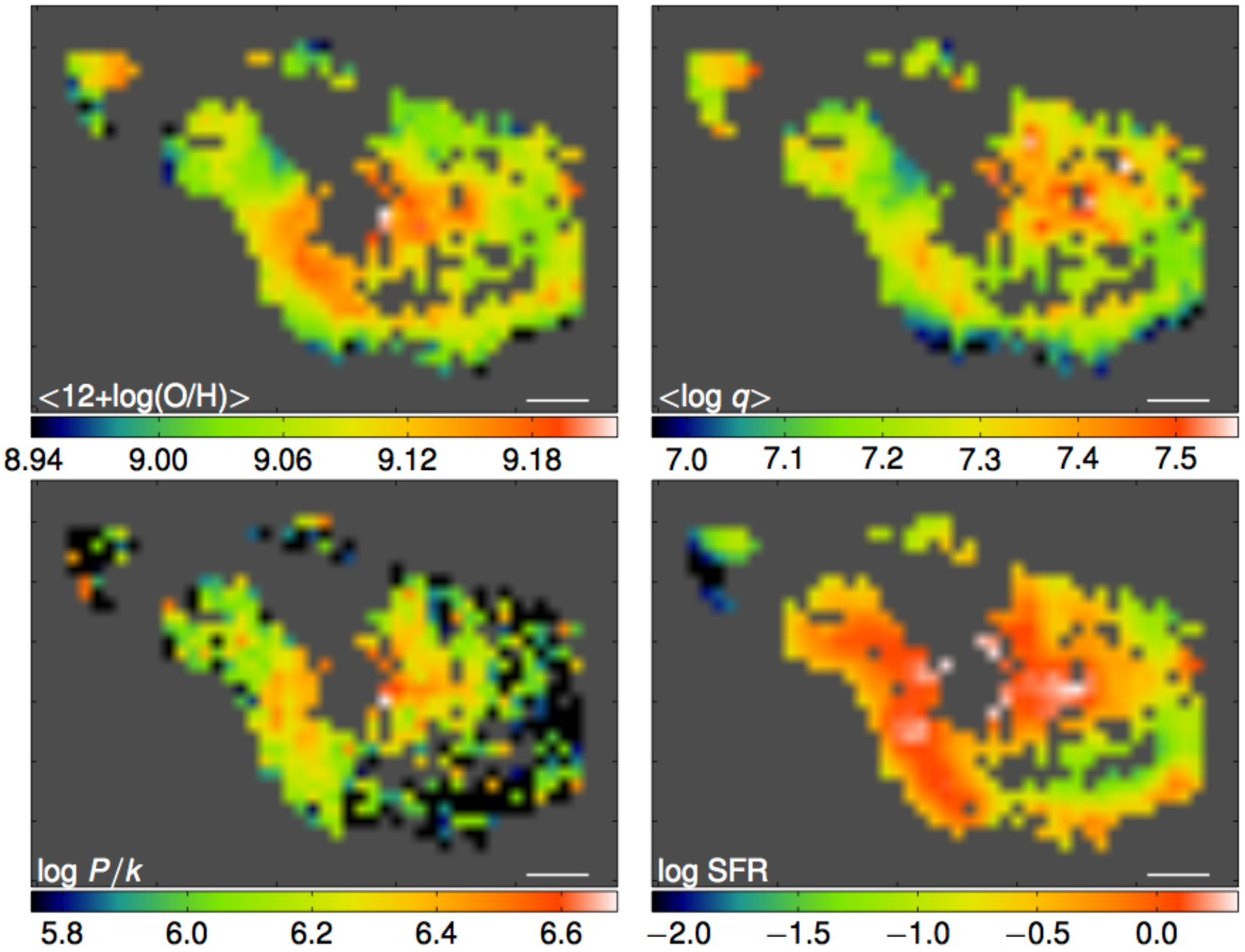}
\caption{As figure \ref{fig_1} but for the galaxy F10257-4339 (NGC 3256). Here (unusually) there is a clear correlation between  $\log(P/k)$, $\log(\rm SFR)$ and  12+$\log$(O/H). However, the observed range of  $\log(q)$ is highly restricted.}\label{fig_4}
\end{figure}

\begin{figure}[ht]
\includegraphics[width=1.0\hsize]{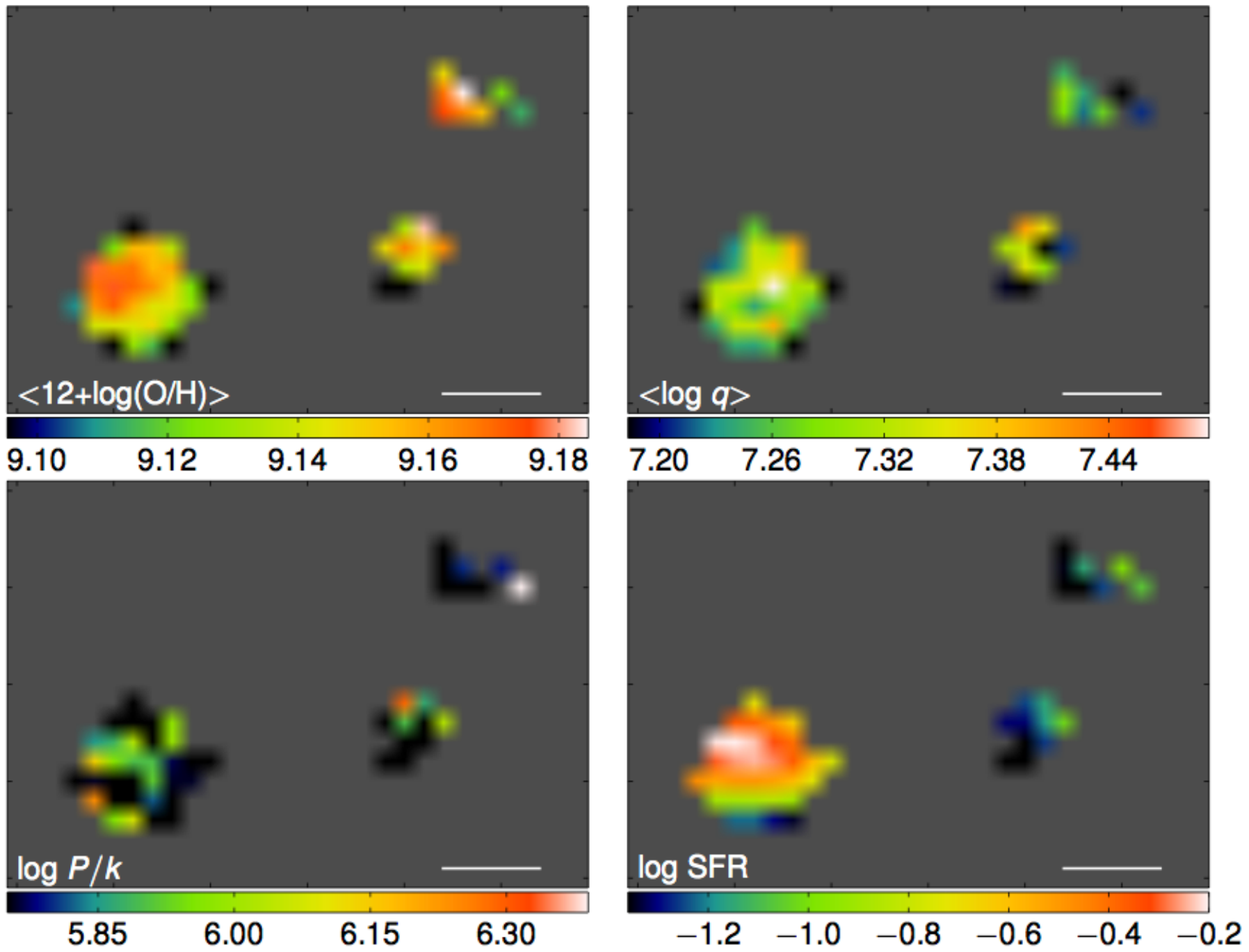}
\caption{As figure \ref{fig_1} but for the object F13373+0105E, a component of Arp240. }\label{fig_5}
\end{figure}

 \begin{figure}[ht]
\includegraphics[width=0.95\hsize]{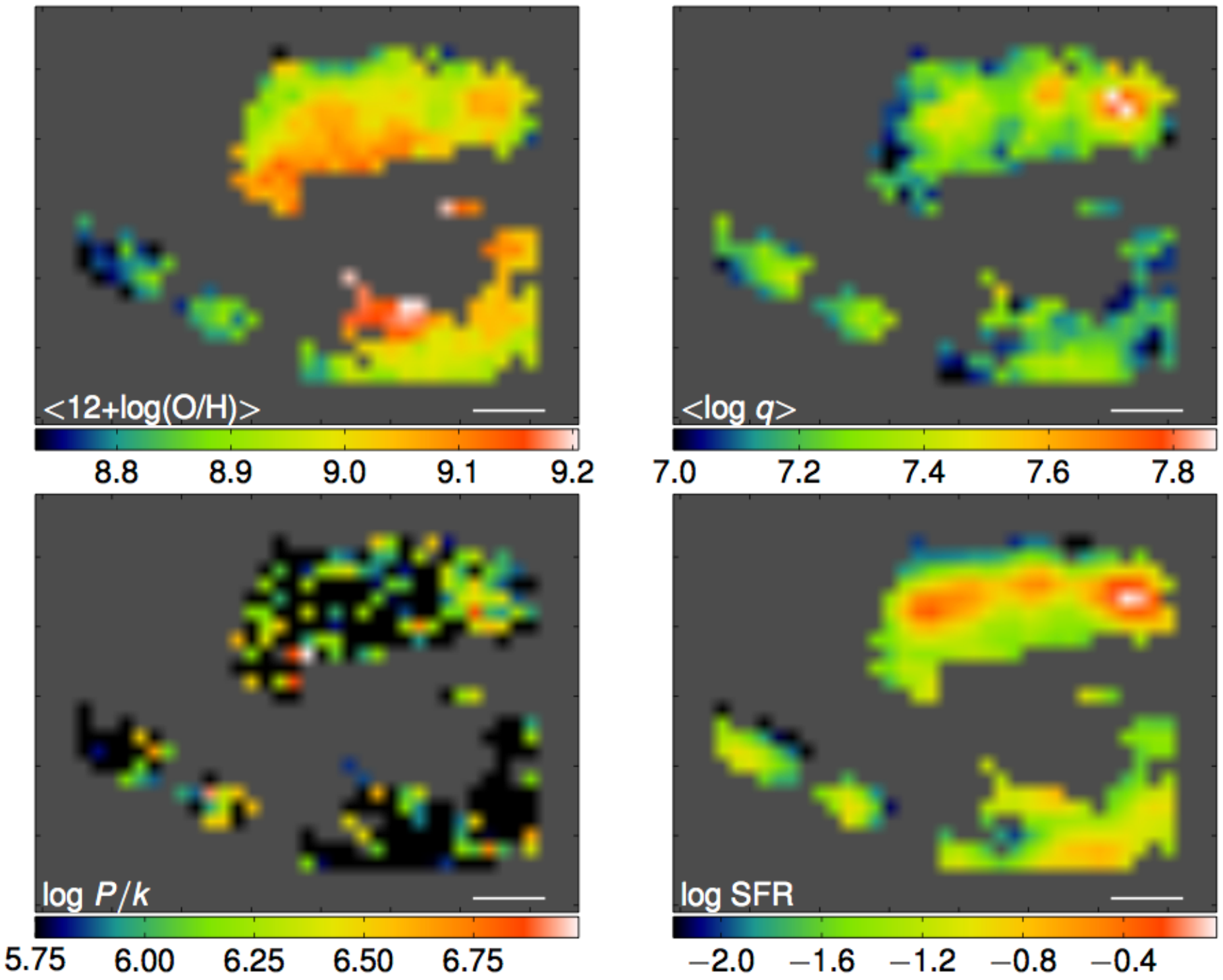}
\caption{As figure \ref{fig_1} but for the object F13373+0105W, a component of Arp240. This is a large barred spiral galaxy, with very little gas near the nucleus (roughly in the centre of the image). Again, the correlation between  $\log(\rm SFR)$ and $\log(q)$ is better than between  $\log(\rm SFR)$ and 12+$\log$(O/H).}\label{fig_6}
\end{figure}

 \begin{figure}[ht]
\includegraphics[width=0.95\hsize]{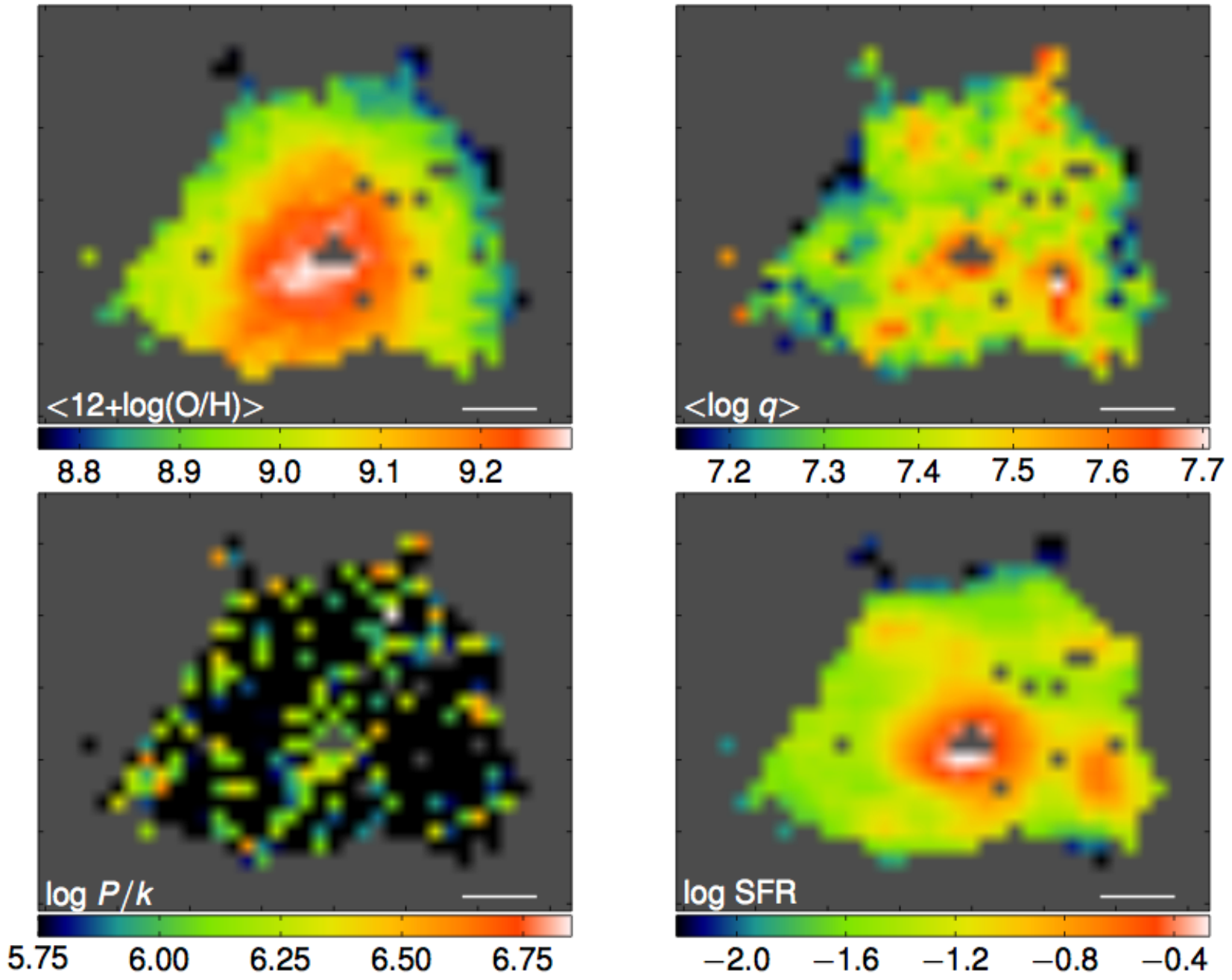}
\caption{As figure \ref{fig_1} but for the object F17222-5953 (ESO 138-G027).}\label{fig_7}
\end{figure}

 \begin{figure}[ht]
\includegraphics[width=1.0\hsize]{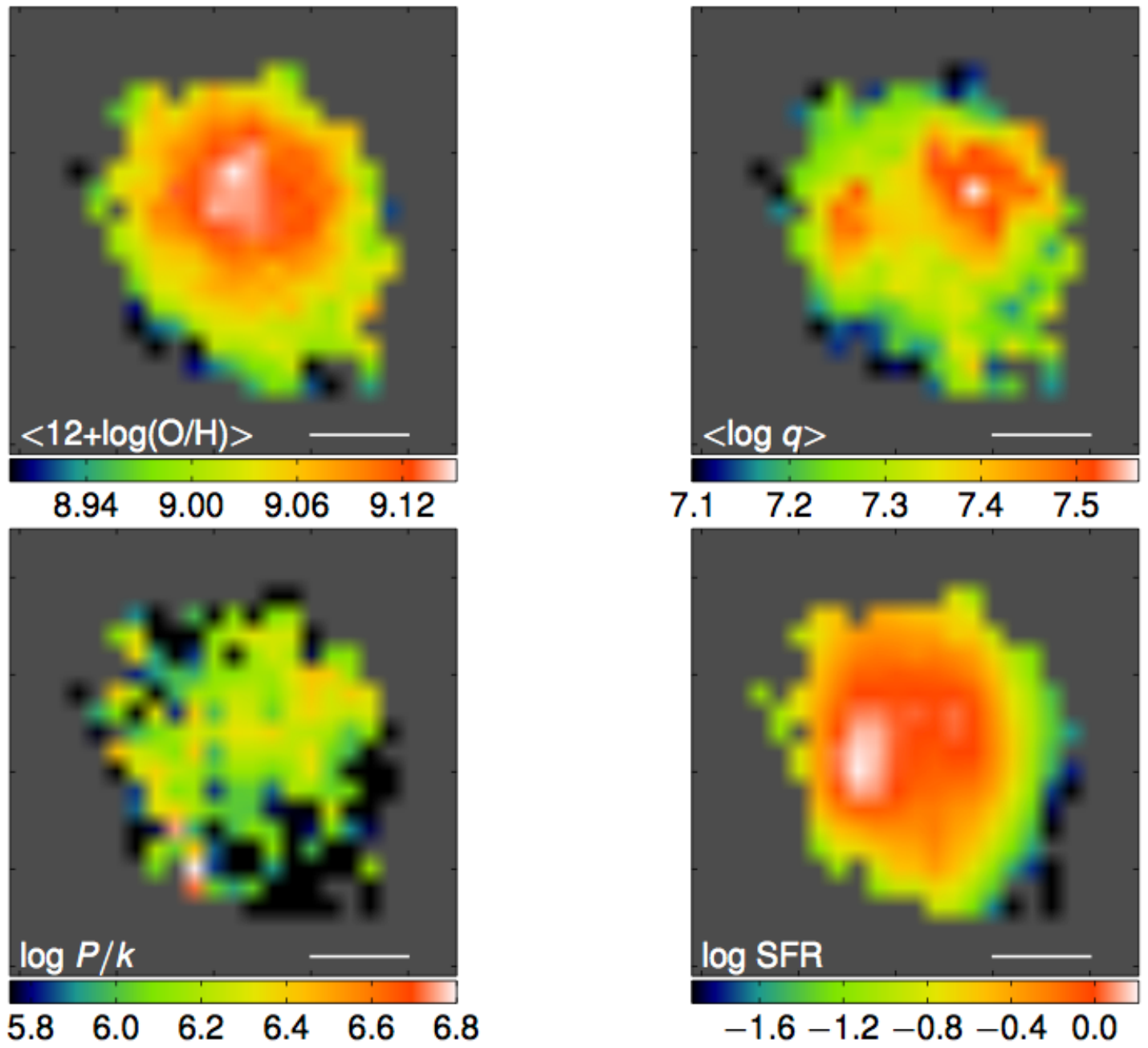}
\caption{As figure \ref{fig_1} but for the object F18093-5744N (IC 4687). This shows clean correlations between $\log(\rm SFR)$, $\log(q)$  and 12+$\log$(O/H).}\label{fig_8}
\end{figure}

 \begin{figure}[ht]
\includegraphics[width=1.0\hsize]{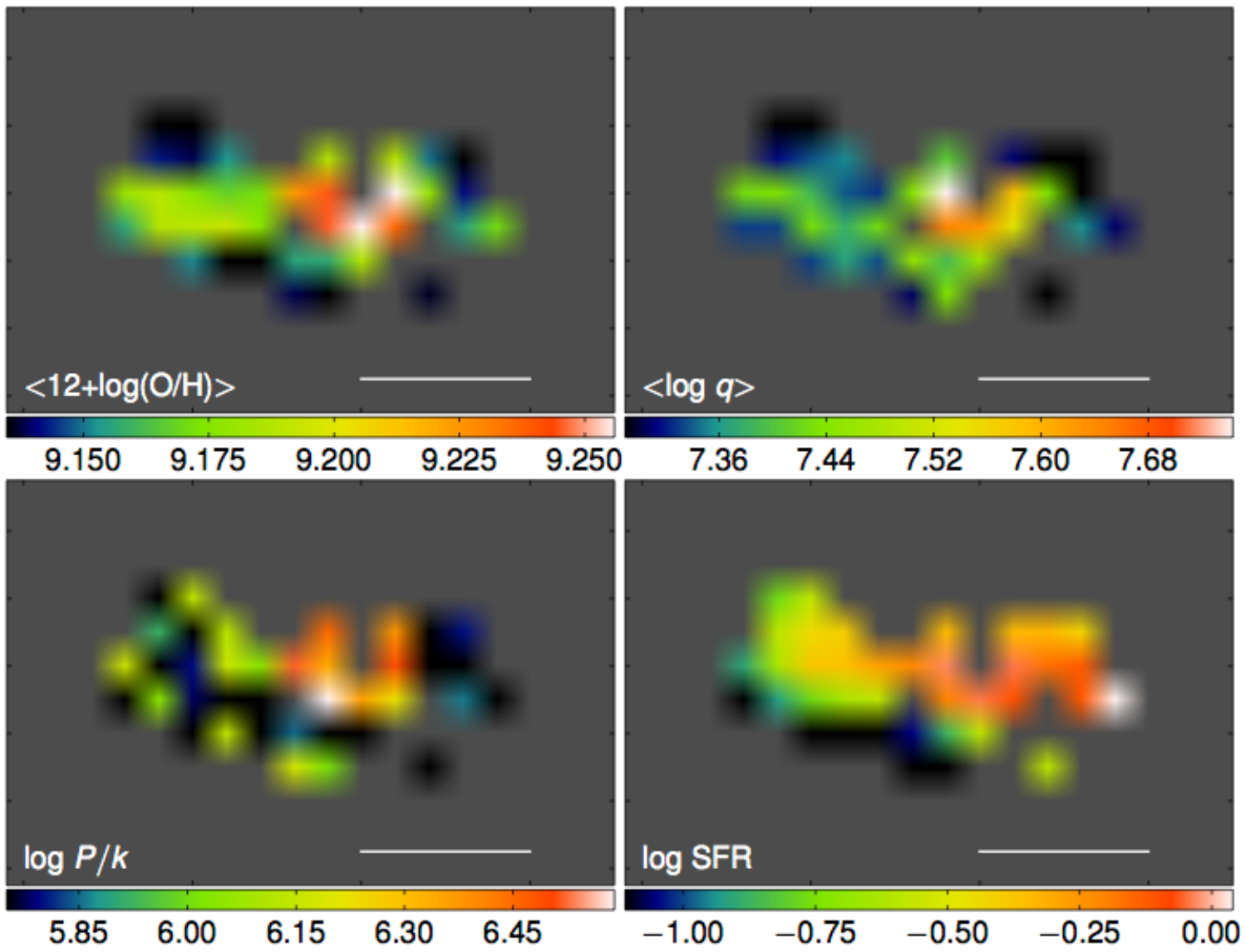}
\caption{As figure \ref{fig_1} but for the object F18093-5744S (IC4689). Despite the relatively small number of spaxels in this object, all quantities plotted seem to be correlated.}\label{fig_9}
\end{figure}

\section{Results}
\subsection{Maps}
The maps for some of the objects are given in Figures \ref{fig_2} to \ref{fig_9}. These include some points that may be contaminated by shock emission. Such points tend to show large dispersions in the  $12+\log(\rm{O/H})$ and $\log(q)$ dervied from the {\sf pyqz} Python module. For the purpose of plotting the correlations between the derived physical parameters, we exclude all points with a standard deviation of greater than 0.1~dex in $12+\log(\rm{O/H})$ and/or greater than 0.15~dex in derived $\log(q)$. The resulting correlations are shown for all galaxies in Figures \ref{fig_10} to \ref{fig_12}.

\subsection{Notes on Individual Objects}
\begin{figure*}[ht]
\includegraphics[width=1.0\hsize]{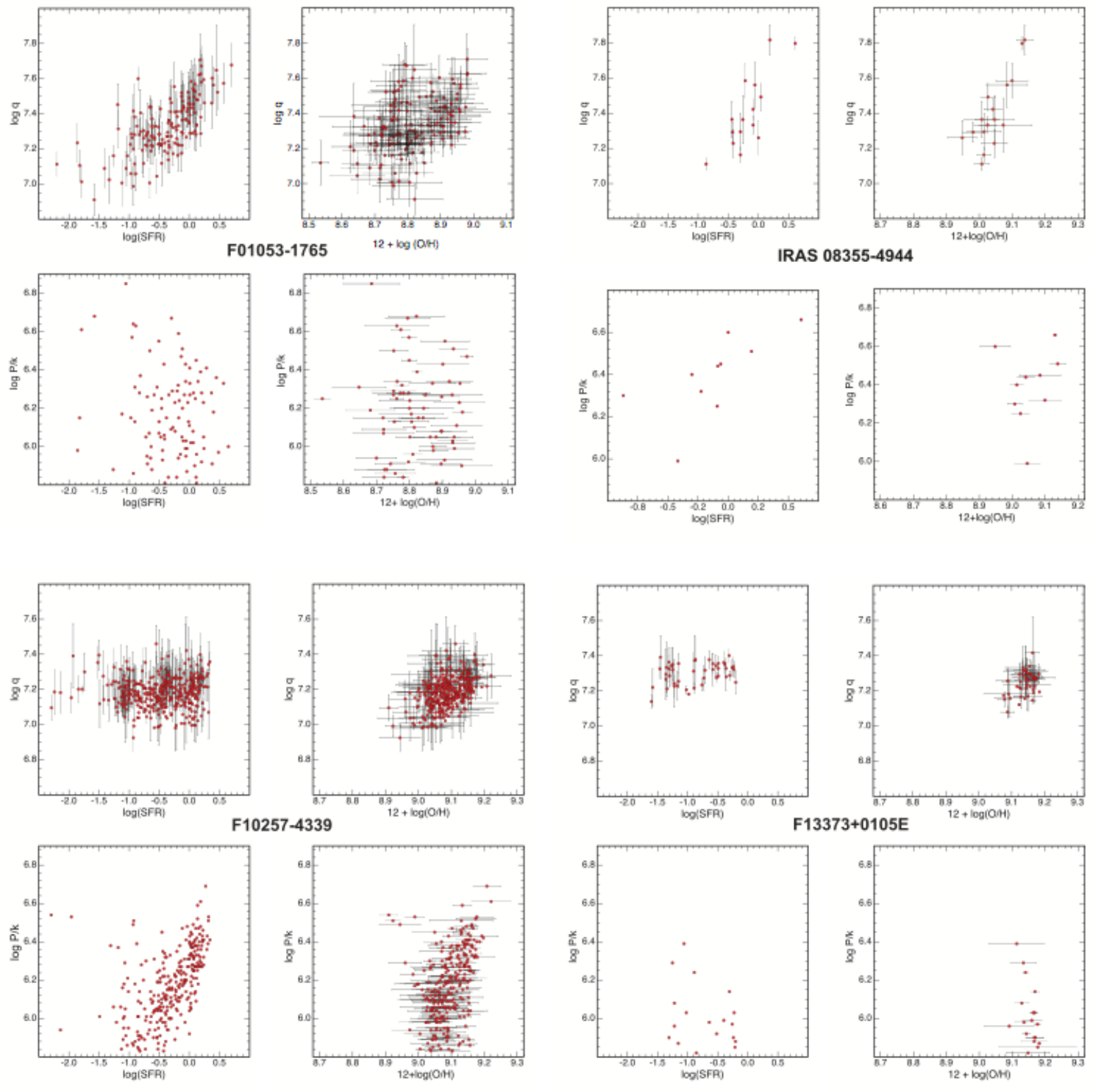}
\caption{Correlations for F01053-1746 (IC1623 A/B), IRAS08355-4944,  F10257-4339 (NGC 3256) and F13373+0105E. For each of these the panels are, top left: the $\log(\rm SFR)$ vs. $\log(q)$ correlation. Here and elsewhere, $\log(\rm SFR)$ should be read as the surface rate of star formation $\log( \Sigma_{\rm SFR})$, measured in units of $M_{\odot}$yr$^{-1}$kpc$^{-2}$. Top right: the  12+$\log$(O/H) vs. $\log(q)$ correlation. Bottom left: $\log(\rm SFR)$ vs. $\log(P/k)$ correlation. Bottom right:  the  12+$\log$(O/H) vs.  $\log(P/k)$ correlation.}\label{fig_10}
\end{figure*}

\noindent {\bf F01053-1746 or IC 1623 A/B:} This is shown in Figure \ref{fig_2}. It is a closely interacting system in the final stages of merging. It is kinematically complex, and shows clear evidence of shock-excited regions as described by \citet{Rich11}. In this, as in other galaxies, we have removed those spaxels which are obviously shock-excited on the basis of the earlier analysis. The eastern member of the pair (the spaxels on the upper left) is clearly the more metal-rich of the two -- a fact which was not evident in the earlier analysis by  \citet{Rich12}. However the abundance range is rather restricted, and intrinsic errors in the earlier strong-line techniques used by \citet{Rich12} (the \citet{Pettini04} method (PP04) scaled to \citet{Kewley02} (KD02) as well as the KD02 diagnostics) would tend to obscure the differences in the oxygen abundance. The separation in metallicity between the two galaxies is clearly evident in the $12+\log$(O/H) vs. $\log(q)$ correlation (see Figure \ref{fig_10}). The western galaxy is the more violently star forming of the two, and it also characterised by the highest ionisation parameters. Overall there is a strong correlation between  $\log(\rm SFR)$ and $\log(q)$. Little can be learnt from the correlations with \HII region pressure.

\noindent {\bf IRAS 08355-4944:} This galaxy (Figure \ref{fig_3}) is a post-merger with long remnant tidal tails. Unfortunately there are too few spaxels to make a meaningful map, but the galaxy shows clear correlations between  $\log(\rm SFR)$ and both $\log(P/k)$ and $\log(q)$. The ionisation parameter is also correlated with  $12+\log$(O/H).

\noindent {\bf F10257-4339 or NGC 3256:} This high-abundance nearby system shown in Figure \ref{fig_4} is another advanced merger displaying widespread evidence of shock-excited gas \citep{Rich11}. The measured ionisation parameters show little evidence of variability across the system, being nearly all in the range $7.4 > \log{q} >7.0$. For this galaxy (Figure \ref{fig_10}) there is a remarkably tight correlation between $\log(\rm SFR)$ and $\log(P/k)$, and also a marked correlation between  $12+\log$(O/H) and  $\log(P/k)$. The weak abundance gradient found here is in agreement with that found by \citet{Rich11}.

\begin{figure*}[ht]
\includegraphics[width=1.0\hsize]{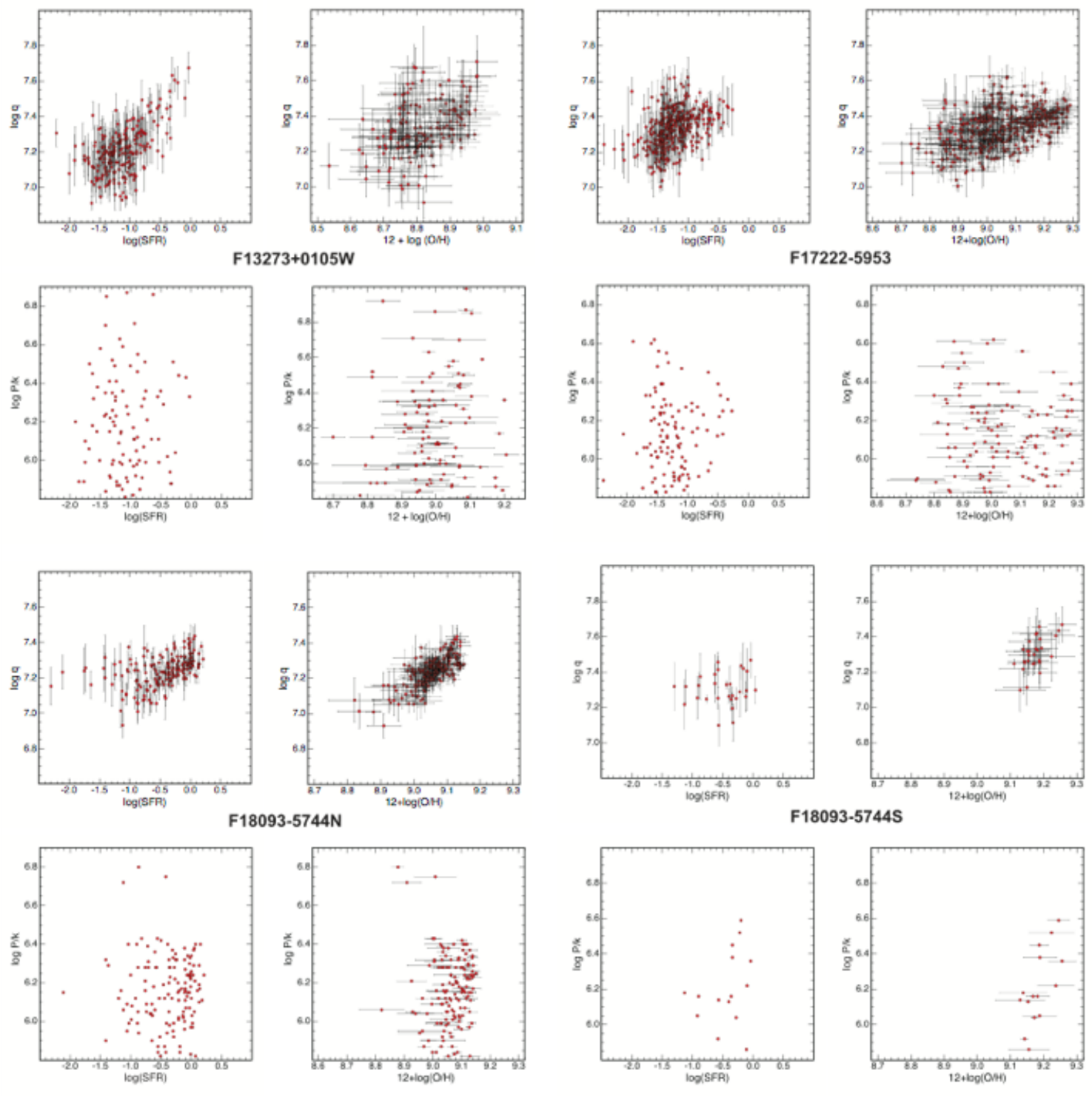}
\caption{As figure \ref{fig_9} but for F13373+0105W, F17222-5953 (ESO 138-G027), F18093-5744N  (IC 4687)and F18093-5744S  (IC 4689). }\label{fig_11}
\end{figure*}

\begin{figure*}[ht]
\includegraphics[width=1.0\hsize]{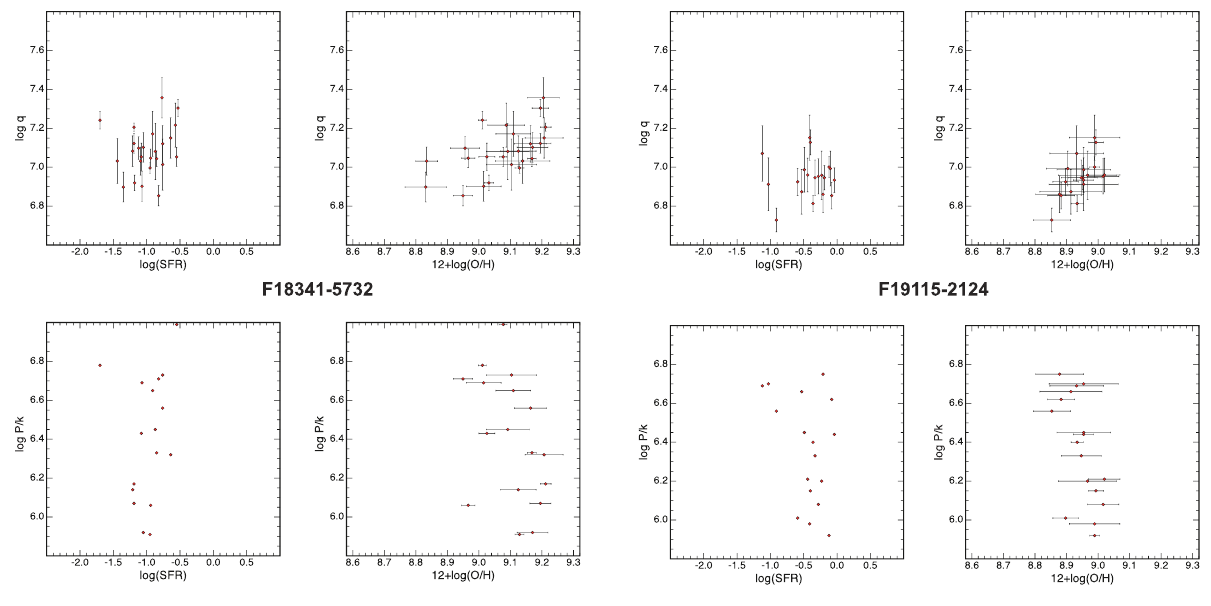}
\caption{As figure \ref{fig_9} but for F18341-5732  (IC 4734)and for F19115-2124  (ESO 593-IG008). }\label{fig_12}
\end{figure*}

\noindent {\bf F13373+0105 E/W or Arp 240:} The eastern component of this system shown in Figure \ref{fig_5} is a tidally distorted spiral (see the HST image in  \citet{Rich11}). Mostly the \HII gas is too faint to support the spaxel analysis, so the map simply picks out the regions of greatest star formation. These show an approximately constant  $\log(q)$ as a function of $\log(\rm SFR)$ (Figure \ref{fig_9}). The western component shown in Figure \ref{fig_6} is much more active \citep{Rich12}, and displays a strong abundance gradient. Our two analyses agree both in the size and the range of this gradient. This object is similar to F01053-1746 in that it displays a strong correlation between  $\log(\rm SFR)$ and $\log(q)$ and a bifurcation in the $12+\log$(O/H) vs. $\log(q)$ correlation (Figure \ref{fig_11}).

\noindent {\bf F17222-5953 or ESO 138-G027:} This system shown in Figure \ref{fig_7} is a typical, non-interacting, strongly star bursting spiral galaxy. It shows a close correlation between  $\log(\rm SFR)$ and $12+\log$(O/H), both of which show a clean radial gradient (Figure \ref{fig_11}). There are positive correlations of both these quantities with $\log(q)$, with again a suggestion of bifurcation in these correlations, caused by the enhanced region of star formation seen in the lower left of Figure \ref{fig_7}, corresponding to a region north of the nucleus on the sky.

\noindent {\bf F18093-5744 N/S or IC 4687/4689:} These two galaxies are in a close triplet system. IC 4687, shown in Figure \ref{fig_8}, is a close merger with the less massive starburst IC 4686. The smaller galaxy IC 4689 (F18093-5744S), Figure \ref{fig_9}, interacts with this pair. IC 4687 (Figure \ref{fig_8} and Figure \ref{fig_10}) displays a strong and clean positive correlation between  both $\log(\rm SFR)$ vs. $\log(q)$ and $12+\log$(O/H) vs. $\log(q)$.  IC 4689 (see Figure \ref{fig_8} and Figure \ref{fig_11}) shows a correlation in $12+\log$(O/H) vs. $\log(q)$, but any correlation with $\log(\rm SFR)$ (if present) is very weak.

\subsection{Global Correlations}
Many of the objects observed show a positive correlation between both $\log(\rm SFR)$ vs. $\log(q)$ and $12+\log$(O/H) vs. $\log(q)$. This is in contrast to observations of \HII regions in general; see the data presented in \citet{Dopita13}, for which there is little evidence for any variation in  $\log(q)$ with $12+\log$(O/H). Nearly all \HII regions fall in a narrow range $7.0 < \log(q) < 7.5$. The outliers tend to be found in lower abundance \HII regions, and these have higher, not lower ionisation parameters. It is clear that the starburst galaxies differ in a fundamental manner from normal galaxies, and that these positive correlations of $\log(\rm SFR)$ vs. $\log(q)$ and $12+\log$(O/H) vs. $\log(q)$ demand some theoretical explanation. 

\begin{figure*}[ht]
\includegraphics[width=1.0\hsize]{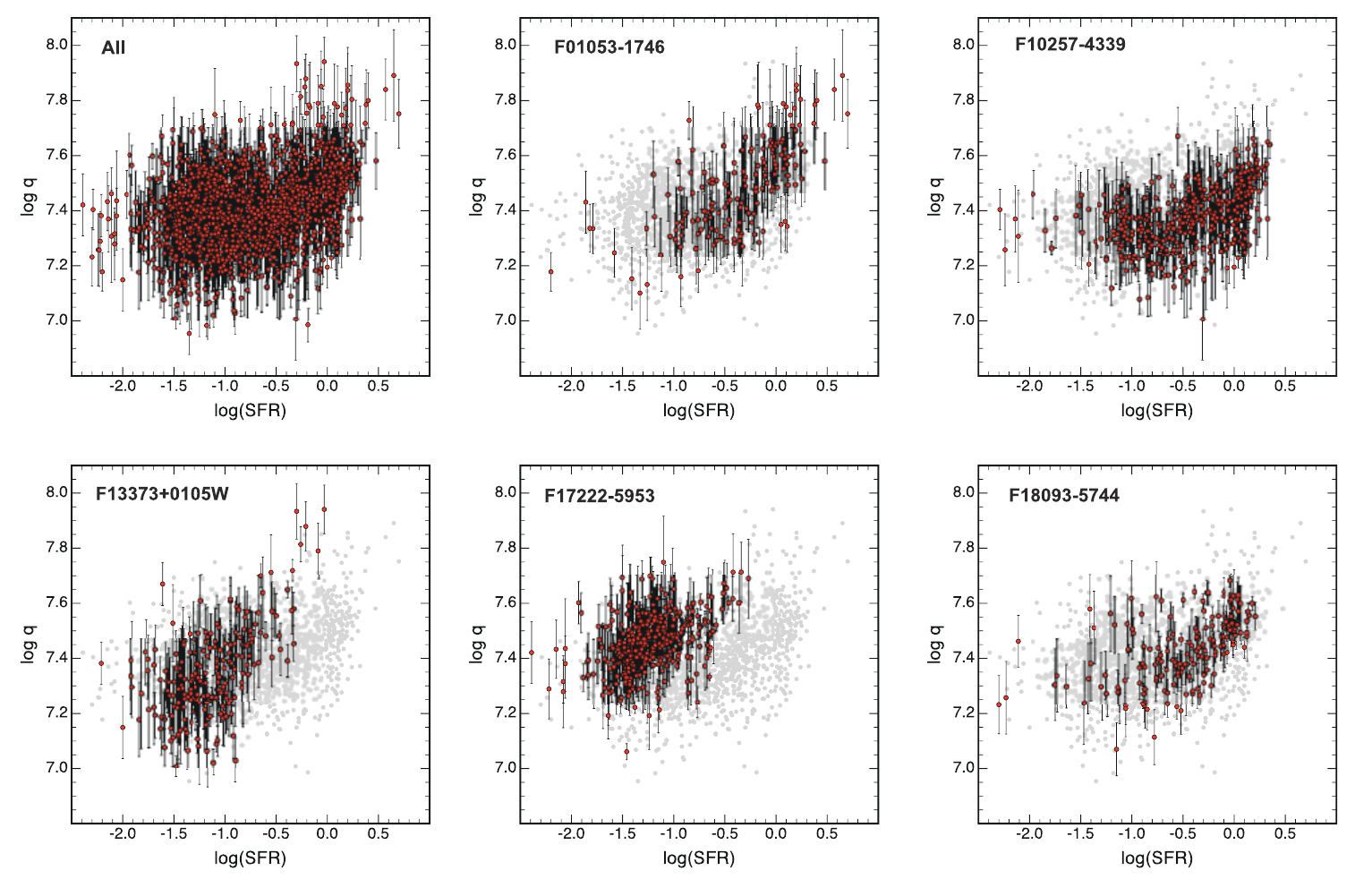}
\caption{Global $\log({\rm SFR})$ vs. $\log q $ correlations corrected for the local pressure. The ensemble of the points for all six galaxies are plotted in pale grey on the background. Above some threshold of star formation rate, which seems to vary galaxy to galaxy, all galaxies display  correlation with a positive slope averaging at $q \propto {\rm SFR}^{0.34}$.}\label{fig_13}
\end{figure*}

\begin{figure}[ht]
\includegraphics[width=1.0\hsize]{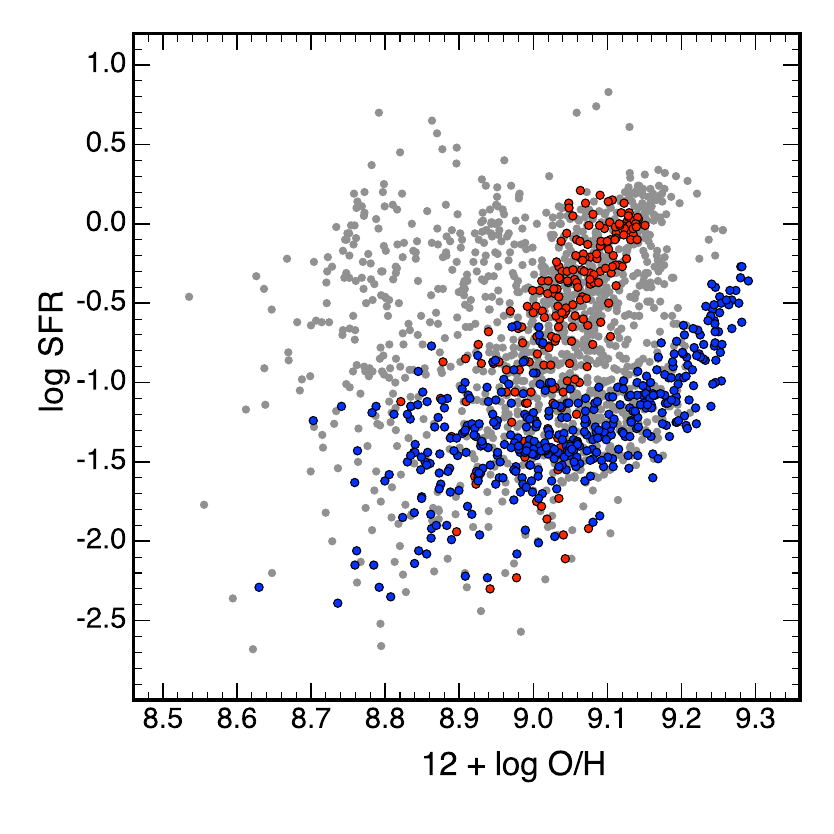}
\caption{The global correlation between $12+\log$(O/H) and $\log$(SFR). For the sample as a whole, we find little evidence that these are correlated, but for individual objects such as F17222-5953 (blue points) or F18093-5744 N (red points) there is a strong correlation. These galaxies display strong abundance gradients, as well as increasing star formation rates towards their centres, presumably associated with higher surface densities of gas. }\label{fig_14}
\end{figure}

Several of the objects shown in Figure  \ref{fig_10}) and  \ref{fig_11} show both positive correlations not only between between  $\log(q)$ and $\log$(SFR) but also between $\log(q)$ and $12+\log$(O/H). Do these then drive a correlation between $12+\log$(O/H) and $\log$(SFR)? We test this in Figure  \ref{fig_14}

\section{Interpretation of Results}

The possible explanations for a systemic increase in the ionisation parameter in regions of high star formation density are as follows:
\begin{enumerate}
\item{ The shape of the stellar Initial Mass Function (IMF) is changing with  $\log({\rm SFR})$. The harder radiation field produced by a flatter IMF could then masquerade as if it were a higher ionisation parameter.}
\item{ The mean cluster mass $M_{\rm cl}$ systematically increases  with  $\log({\rm SFR})$. The collective effects of a massive cluster speed up the expansion of the surrounding \HII region, and produce a higher ionisation parameter at a given cluster age.}
\item{ The \HII regions may overlap each other.}
\item{ The structure of the \HII\ regions is dependent on the physical environment of the starburst region. }
\end{enumerate}
We will now explore each of these possibilities in turn.

\subsection{Effect of IMF slope}
In order to check the effect of an environment dependent IMF slope, we have run a new set of Starburst 99 v6.0.2 \citep{SB99} models with continuous star formation extending over 3Myr  - typical of the distribution of ages in massive young clusters exciting the most luminous \HII regions \citep{Beccari10, DeMarchi11}. These models have different slopes of the upper ($0.5 - 120$ M$_{\odot}$) IMF; -2.6, -2.2, -1.8 and -1.4. 
To be as closely comparable to the models used by \citet{Dopita13} as possible, we used the Geneva tracks for metallicity Z=0.04, and the \citet{Lejeune97} model atmospheres. For stars with strong winds we switch to the  \citet{Schmutz92} extended model atmospheres using the prescription of \citet{Leitherer95}.

\begin{figure}[ht]
\includegraphics[width=1.0\hsize]{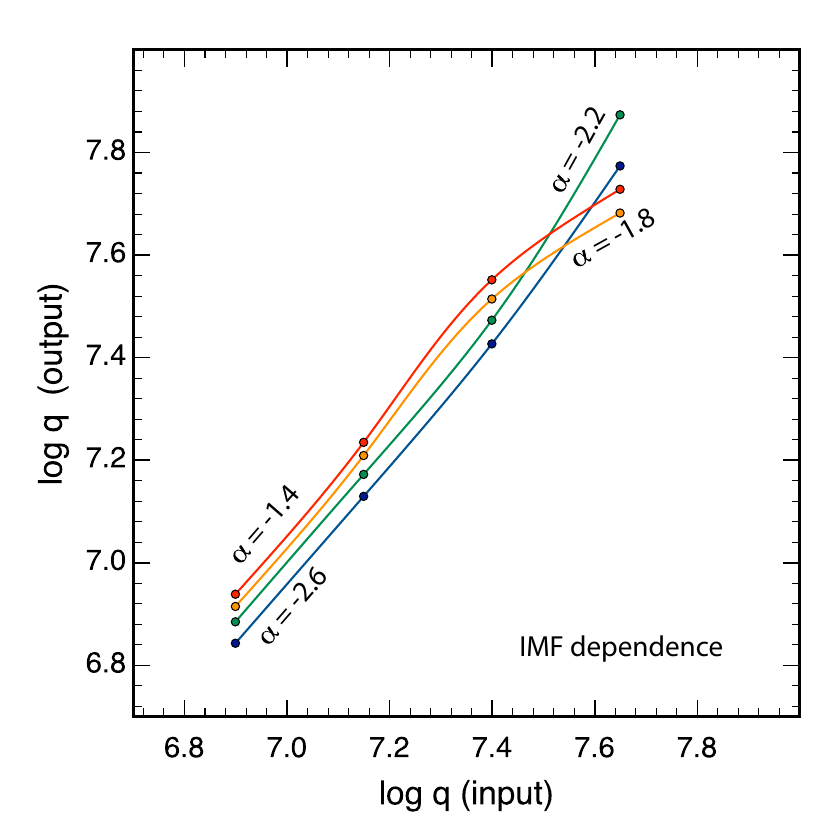}
\caption{The $\log q $ input into the \emph{Mappings IV} code vs. the output  $\log q $ as determined by the {\sf pyqz} software \citep{Dopita13}. Four different input IMF were used with upper IMF slopes of $\alpha = -2.6, -2.2, -1.8$ and -$1.4$, respectively. The IMF slope has only a small effect on $q$, and IMF variations alone cannot explain the correlation between $\log(SFR)$ and $\log(q)$.}\label{fig_15}
\end{figure}

These atmospheres were input into the Mappings IV code \citep{Dopita13}. We ran isobaric models with a pressure of $\log (P/k) = 10^6$cm$^{-3}$K, typical of our starbursts, a 3 times solar abundance set (oxygen abundance $12+ \log({\rm O/H}) = 9.17$) and  input ionisation parameters of $q =6.9, 7.15, 7.4$ and 7.65. The resultant spectra were then fed though through the {\sf pyqz} software  \citep{Dopita13} to solve for the apparent (output) ionisation parameter and the chemical abundance. This procedure recovered the input abundances within 0.05 dex, but the output $\log q$ was somewhat different from the input value. The result is shown in Figure \ref{fig_15}.

The effect of the IMF slope is rather small. Of course we have controlled against first order effects by choosing the ionisation parameter at the inner edge of the nebula. Any residual effect is due to changes in the shape of the ionising radiation field. The somewhat harder radiation field of the flatter IMF does indeed raise the derived $\log q$, but only by about 0.1 dex at most. However, this effect tends to reverse at the highest input values of $\log q$ due to the different temperature structure of the nebula, which shows steep temperature gradients through the models at the high chemical abundances characterising our model \HII regions. We conclude that IMF variations which are a function of the local star formation rate cannot explain the size of our derived correlation between $\log q$ and $\log ({\rm SFR})$.

\subsection{Spherical \HII regions: Effect of  cluster mass}
\citet{Dopita06} developed a simple theory describing the development of a spherical \HII region trapped between the outer boundary of an expanding bubble powered by the wind from the central cluster, and the shocked stellar wind itself. In this idealised model, the pressure in the \HII region is supplied by the pressure of the shocked stellar wind. The central cluster provides the EUV photons which ionise the nebula, while the temporal evolution of the bubble (determined by the mechanical energy luminosity of the central cluster, and the density of the interstellar medium) fixes the radius. In this theory, the ionisation parameter at the contact discontinuity in the photoionised plasma depends simply on the instantaneous properties of the exciting cluster stars;
\begin{equation}
q(t) \propto \delta(t)^{3/2}S_*(t)/L_{\rm mech}(t), \label{eqn_3}
\end{equation}
where $ \delta(t)$ is the instantaneous ratio of the density in the \HII region to the density in the interstellar medium (ISM) surrounding the \HII region, $S_*(t)$ is the instantaneous flux of ionising photons produced by the central cluster, and $L_{\rm mech}(t)$ is the instantaneous mechanical energy flux from the cluster. Clearly, the ratio of two main terms in Equation \ref{eqn_3},  $S_*(t)/L_{\rm mech}(t)$,  is unique for a given slope of the upper IMF, $\alpha_3$ , and for a given stellar metallicity. Provided that these variables are fixed, the $\delta(t)$ term provides for a weak coupling between $q(t)$,  the mean cluster mass,  $M_{\rm cl}$, and the  pressure in the interstellar medium, $P_0$; $q \propto \left( M_{\rm cl}/P_0 \right)^{1/5}$. This scaling is evident in Figure 5 of  \citet{Dopita06}.

\citet{Dopita06} also showed that the ionisation parameter is rather sensitive to the chemical abundance, $Z$. At high abundance the stellar wind has higher opacity, and absorbs ionising photons before they escape the extended atmosphere, and the mechanical luminosity is higher leading to greater compression of the \HII shell. Both of these effects lower $q$ leading to $q \propto Z^{-0.8}$, approximately over the whole range of abundance, although the abundance effect is much weaker at the high chemical abundances characterised by these objects. Note however that we have found a \emph{positive} correlation between $q$ and  $12+\log$(O/H) - exactly the opposite of what is predicted by theory.

The pressure effect on $q$ predicted in the theory can be readily removed, since we have already measured the pressure in most spaxels. We take $\log(p/k) = 5$ as the reference pressure, since all of our photoionisation models use this as the default, and we correct the measured $q$ by the amount given by the theory; a factor $[(P/k)/10^5{\rm cm^{-3}K}]^{1/5}$. Since, in general, the measured pressure in any spaxel is higher than $\log(p/k) = 5$ , the effect of correcting for pressure in this way serves to increase the effective $q$. This tends to reverse the correction accounting for the change in the line ratios with pressure ( see Section 3). The result is shown in Figure \ref{fig_13} for the six objects for which many spaxels have been measured. For each of these, we also show in pale the distribution of all measured points. In general, we now have a \emph{positive} correlation between $q$ and star formation for all galaxies, although there seems to be a threshold in star formation rate (which differs galaxy to galaxy) above which the correlation becomes strong.  Where the strong correlation exists:
\begin{equation}
q \propto {\rm SFR}^{0.34\pm0.08}.
\end{equation}
There is some evidence of a threshold value for $\log q \sim 7.2 - 7.4$ below which this correlation disappears. Coincidentally (or not?) this is the value of $\log q$ which characterises most individual extragalactic \HII regions \citep{Dopita13}.

If we adopt the mass-loss bubble driven model of the evolution of \HII regions from \citet{Dopita06}, then the relation between cluster mass and ionisation parameter is $q \propto M_{cl}^{1/5}$. This is a very weak dependence and implies that large variations of cluster mass as a function of star formation density are needed to drive the observed variation in  $\log q$. We derived above that $q \propto \Sigma_{\rm SFR}^{0.34\pm0.08}$, where the star formation rate is now written as $\Sigma_{\rm SFR}$ to emphasise that we are talking about surface rates of star formation. If the simple spherical evolution model is applicable, and if the star formation rate \emph{vs.} $q$ correlation is entirely driven by a change in mean cluster mass, these relations together imply that $M_{cl} \propto \Sigma_{\rm SFR}^{1.7\pm0.4}$.

 If, for example, the mean cluster mass at a star formation rate of 0.1 \Msunpyr kpc$^{-2}$ is $10^4 \Msun$, then for a region with 3.0 \Msunpyr kpc$^{-2}$, the mean cluster mass would rise to $3\times 10^6 \Msun$. Such values are not forbidden according to the arguments of the previous sub-section. Indeed, in the case of one of our objects, F10257-4339 or NGC~3256, a survey of its clusters by \citet{Trancho07} finds that they have masses in the range $0.2-4\times 10^{6} \Msun$. In other galaxies even more massive clusters are found. For example, in NGC 7252 and NGC 1316 clusters have been found with masses significantly above $10^7 \Msun$ \citep{Schweizer98,Maraston04,Bastian06}.

The idea that cluster masses scale up with star formation rates has some support from a number of observational and theoretical papers. Most directly, the mass function of clusters has been found by a number of authors to follow a \citet{Schechter76} distribution,
\begin{equation}
dN/dM = A M^{-\beta} \exp[-M/M_*].
\end{equation}
where $\beta \sim 2$, and $M_*$ is the cut-off mass. For ordinary disk galaxies, $M_* \approx 2\times 10^5 \Msun$ \citep{Gieles06,Larsen09}, while for luminous IR galaxies, \citet{Bastian08} finds  $M_* \approx 10^6\Msun$;  \citep{Portegies10}. However, other studies such as that by \citet{Chandar10} find no evidence for an upper mass cutoff. 

This idea was recently developed on a theoretical basis by \citet{Powell13}. They studied merger-induced star formation with 5 pc resolution adaptive mesh refinement simulations of low-redshift equal-mass mergers with randomly chosen orbital parameters. They found an enhanced mass fraction of very dense gas that appears as the gas density probability density function evolves during the merger, which opens up the possibility that such dense regions may give rise to more massive clusters. \citet{Powell13} argue that this dense gas may reveal itself in terms of the enhanced HCN(1-0)/CO(1-0) ratios which are observed in ULIRGs \citep{Juneau09}. Simulations at even higher resolution by \citet{Hopkins13} lead to similar conclusions. They find that the final starburst is dominated by in situ star formation, fuelled by gas which flows inwards due to global torques. High gas density results in massive giant molecular clouds, and rapid star formation leading to the formation of super star clusters with masses of up to $10^8 \Msun$.

Our simple spherical model of \HII region evolution implies that the surface number density of clusters has to decrease as the surface density of gas increases. Crudely speaking,  $\Sigma_{\rm SFR} \propto \Sigma_{cl} M_{cl}$, where $\Sigma_{cl}$ is the surface number density of young clusters. Given that the \citet{Kennicutt98} star formation law has $\Sigma_{\rm SFR} \propto \Sigma_g^{1.4\pm0.1}$, where $\Sigma_g$ is the gas surface density, and that $M_{cl} \propto \Sigma_{\rm SFR}^{1.7\pm0.4}$, then we have that $\Sigma_{cl} \propto \Sigma_g^{-1}$, approximately. Thus dense regions should have fewer clusters, but individually they would be much more massive. As pointed out by our referee, this is in direct contradiction to observation, since the density of clusters does not decrease in this manner. Galaxies with the highest gas densities (e.g., NGC 3256, Arp 220, and to a lesser extent the Antennae galaxies) also have more clusters, and the clusters are strongly positively correlated with the gas with individual galaxies \citet{Zhang01}.  So higher gas surface densities lead to more clusters.  This result also comes out of the analytic calculations of  \citet{Kruijssen12}, although in his model, the clusters are also destroyed faster due to the high gas densities (within a few Myrs) so there is a ``sweet spot'' for surviving clusters. 

In conclusion, while cluster masses may indeed vary as a function of star formation density, this effect is not by itself sufficient to drive the observed slope of the $\log q$ : $\log$(SFR) relation. We must therefore appeal to geometrical effects. These include of either the effect of overlapping \HII regions, or more realistic non-spherical geometries of the \HII regions.

\subsection{ \HII region overlap}
One possible way of changing the effective $\log q $ in the \HII regions is to make them interfere spatially with each other, which would put ionised gas closer to the ionising source. However, it is not necessarily the case that this would enhance the $\log q $ in the ionised gas, since the external pressure of the ionised gas is determined by the pressure of the OB-star stellar wind of the exciting cluster. A reduction in the effective radius of the \HII region will lead not only to an increase in the photon flux, but also to a corresponding increase in the confining pressure, which would tend to keep  $\log q $ fixed. The degree to which this statement is true will depend on the position of the inner shock, where the kinetic energy of the stellar wind is thermalised, and whether the ionised material lies within or without the inner shock. Putting these issues of detail aside, let us first consider whether in principle the \HII regions in violently star-bursting regions can become so closely packed that they can run into each other. 

The peak star formation in our objects approaches 3.0 \Msunpyr kpc$^{-2}$. If all of this were to occur in a single cluster over the $\sim 3$~Myr that the cluster produces a strong UV radiation field, then the maximum cluster mass could reach $M_{cl} \approx 10^{7} \Msun$. Of course, the star formation is much more likely to be distributed amongst a number of smaller clusters. Typical \HII region parameters have $\log(P/k) \sim 6.2$~cm$^{-3}$K and $\log q \sim 7.5$~cm~s$^{-1}$. With these parameters, the \HII region forms a thin ionised shell around the hot shocked stellar wind, and the radius of the \HII region is determined by the value of $\log q$. The radius scales as the square root of the cluster luminosity or cluster mass. From detailed \emph{Mappings IV} photoionisation models:
\begin{equation}
R_{\rm HII} = 68[M_{cl}/10^6 \Msun]^{1/2}~{\rm pc}. \label{radius}
\end{equation}
Thus, the maximum volume of the stellar wind plus \HII region is achieved when all of the star formation is concentrated in a single cluster. This volume is $\sim 4\times10^7$ pc$^3$. However the total volume available is  $10^8 - 10^9$ pc$^3$, depending on whether the available thickness of the gas layer is 100pc or 1.0kpc. We conclude that, at the high pressures characteristic of these starburst regions, interference between the individual \HII regions around clusters is rather unlikely to occur.

\subsection{Non-spherical \HII regions.}
The approximation of spherical \HII region expansion is clearly too simplistic. The interstellar medium of galaxies is turbulent and fractal, and regions of active star formation are constantly shifting to where dense molecular clouds are actively collapsing. Where dense molecular clouds are overrun by the expanding shell of ionised gas, ``elephant trunks'' are formed, each surrounded by an outflowing photo-evaporation zone. In some cases, an embedded \HII region breaks out into a lower density region, forming a ``champagne flow'' \citep{TT79,Arthur07,Hu08}.

In the case of a champagne flow, very high ionisation parameters can occur since the molecular gas remains very close to the ionising source, limited only by the photo-ablation timescale of the molecular cloud. Something of this kind may be occurring in M82.  \citet{Smith06} and \citet{Westmoquette07} used HST imaging and spectroscopic analysis of super star clusters in M82 and found that the  most compact region A1 contains a massive  ($\sim 0.5-1\times10^6 \Msun$) young ($6-7$Myr) and compact ($\sim3.5$pc radius) cluster,  but hosts a HII region around it with a size of only 4.5 pc. This \HII region has an internal pressure of $P/k = (1-2)\times 10^7$ cm$^{-3}$K and with an ionisation parameter $\log q \sim 8.2$ cm s$^{-1}$ ($\log U \sim -2.2$). The high ionisation parameter is observed to extend over a region 500 pc across. Thus, the whole region seems to represent a (more extreme) version of the high -$q$, high - SFR regions studied in this paper. 

We should first note that the measured value of the ionisation parameter for this object corresponds to the limit set by where the radiation pressure (acting mainly on dust) exceeds the gas pressure at the inner edge of the ionised region closest to the exciting stars. In this case radiation pressure serves to compress the ionised gas close to the ionisation front \citep{Dopita02} and dust becomes the main absorber of the ionising photons\citep{Dopita03}. 

The ionised region around the cluster A1 cannot be spherical. Compared with the example given by equation \ref{radius}, the higher pressure alone leads to a 25 times reduction in the Str\"omgren volume, and the absorption of the ionising photons by dust ($\sim 70$\%), makes the total reduction in the Str\"omgren volume a factor of 80. Nonetheless, we would still expect the \HII region to have a radius $\sim 10-15$pc in radius, compared to the 4.5pc observed. Thus, either it is a ``blister'' \HII region at the edge of a massive molecular cloud, or else it contains massive molecular clouds much closer to the central stars than the spherical shell model would predict. Either way, the nebular geometry is distinctly non-spherical.

  \citet{Smith06} and  \citet{Westmoquette07} suggest that the expansion of the \HII region is ``stalled'' by the dense environment. However, the concept of stalling the expansion is dubious, since the internal pressure of the \HII region from ionised plasma, stellar winds and supernova explosions will always exceed the pressure in the ISM, for any reasonable values. The fundamental  question is whether the expansion velocity is supersonic or subsonic with respect to the turbulent velocity of the neutral (or molecular) medium. The possibility that the expansion velocity of the \HII region during the lifetime of the OB stars within it may be less than the characteristic random velocity of the molecular clouds in its vicinity opens up the interesting possibility that the molecular clouds may ``fall'' into the \HII region and become highly ionised (high- $q$) as they pass close to the exciting stars. In this geometry, not only do we have a Kelvin-Helmholtz unstable layer at the surface of the clouds which leads to a turbulent outflow, but in the high- $q$ conditions produced, we may also have radiative acceleration acting on dust, which can lead to high outflow velocities \citep{Dopita02}. Indeed \citet{Westmoquette07}  conclude that (quoting their paper):
 \begin{quotation} 
 Evaporation and /or ablation of material from  interstellar gas clouds caused by the impact of high-energy photons and fast flowing cluster winds produce a highly turbulent layer on the surface of the clouds from which the emission arises.
\end{quotation}
Molecular clouds may also be infalling towards the stars as part of a general accretion flow, provided that the mass of the central cluster and its placental molecular cloud complex is large enough.

\begin{figure}[ht]
\includegraphics[width=1.0\hsize]{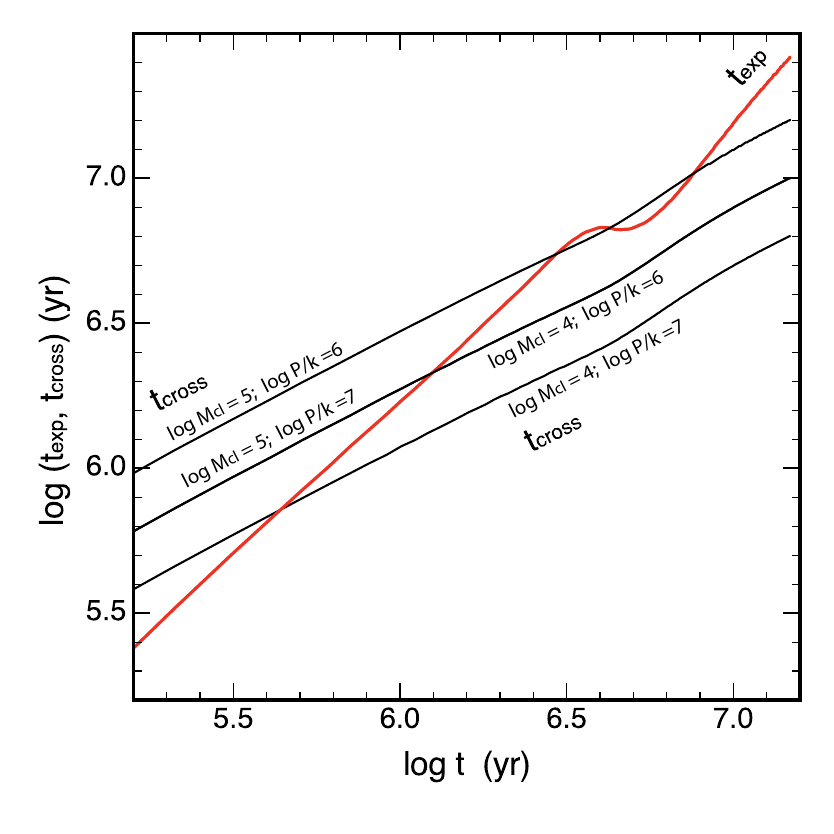}
\caption{The expansion timescale of a (spherical) \HII region compared with the infall timescale of molecular clouds in a turbulent ISM. The infall timescale maybe appreciably smaller than the expansion timescale once the \HII region is older than about $10^6$yr.} \label{fig_16}
\end{figure}

In order to check whether the clouds can cross the \HII region within the lifetime of the OB stars, we have used the mass-loss bubble driven model of the evolution of \HII regions from \citet{Dopita06} to compute the expansion timescale of the bubble, and the infall timescale of a cloud from the edge of the bubble to the centre, assuming a turbulent cloud velocity of 15 km~s$^{-1}$, similar to the velocity dispersions inferred by direct observation \citep{Dickey90} or theory \citep{Joung09}. The results are plotted in Figure \ref{fig_16} for two cluster masses and \HII region pressures. For an appreciable fraction of the lifetime, the infall timescall is less than the expansion timescale. Lower cluster masses, and higher pressures both favour the infall of molecular clouds. Thus, in dense starburst regions, the dynamical evolution of the molecular gas around the \HII region cannot be ignored.

\section{Conclusions}
We conclude that the spaxel analysis of data cubes obtained in luminous infrared galaxies provides useful measures of the local chemical abundance, ionisation parameter, gas pressure and star formation rates.  Positive correlations are found between the ionisation parameter and either the star formation rate or the local chemical abundances. We expect star formation rates and chemical abundances to be correlated, since the inner regions of these galaxies generally have both higher star formation rates and higher chemical abundances than the outskirts. However, the fact of a correlation of these quantities with ionisation parameter is more intriguing. We have corrected for the effect of the local pressure, but this only serves to make the $q$: SFR correlation stronger, as shown in Figure \ref{fig_12}. 

We have examined a number of possible causes for the correlation;
\begin{itemize}
\item{ That initial mass function variation drives the correlations.}
\item{That the mass of the exciting cluster changes systematically with environment.}
\item{That in high star formation regions, overlap of the \HII regions occurs.}
\item{That the geometry of the molecular and ionised gas changes strongly with environment.}
\end{itemize}
We are able to eliminate the first and third of the possibilities listed above. While the mass of the exciting clusters may well become higher as the star formation rate per unit area becomes higher, this effect is insufficient to drive the observed correlation. We conclude that the geometrical effect is the most likely cause, and that the \HII regions in the dense star forming regions contain inclusions of molecular clouds undergoing violent radiation pressure dominated photo-ablation close to the exciting stars, analogous to that described by \citet{Smith06, Westmoquette07} in the nuclear region of M82. In these circumstances, fast radially-directed outflows from the molecular clouds can be driven by radiative pressure. This property may well be the root cause of the correlation between the turbulent width of the H$\alpha$ line and the H$\alpha$ luminosity found by many authors e.g. \citet{Green10,Wisnioski12} and \citet{Swinbank12}.

The model of radiation pressure dominated photo-ablation of dense molecular clouds embedded deep within \HII regions in high specific star formation regions should give rise to consequences amenable to direct observation. First, the clouds themselves should be visible in high-resolution H$\alpha$ images taken with the Hubble Space Telescope. A good example of such an environment is M83, where the central pressure and star formation rate achieve values comparable with the cores of the ULIRGs studied here; \emph{c.f} \citet{Dopita10b}. In addition, in these objects high-resolution integral field spectroscopy (either in the IR or at optical wavelengths) should reveal local line broadening due to the high-velocity comet-like ablation flows expected at the edges of the dense molecular inclusions. A more extreme example of such a flow has been observed with HST in the nearby Seyfert 2 galaxy, NGC~1068 \citep{Cecil02}. Here the ouflow velocities range up to thousands of km~s$^{-1}$, while for the photoablation flows we are considering velocities would be much lower; $\sim 100$km~s$^{-1}$.

\begin{acknowledgments}
Mike Dopita and Lisa Kewley acknowledge the support of the Australian Research Council (ARC) through their Discovery project DP130103925  This work was also funded by the Deanship of Scientific Research (DSR), King Abdulaziz University, under grant No. (5-130/1433 HiCi). The authors acknowledge this financial support from KAU. We also thank the anonymous referee for insightful comments which led us to re-examine our fundamental assumptions. The paper has been much improved as a result of this valuable input.
\end{acknowledgments}

\end{document}